\documentclass{phbauth}

 \usepackage[dvips]{graphicx}
\usepackage{amssymb}

\begin{document}

\begin{frontmatter}



\title{The effect of charge fluctuation on a normal-superconducting-normal 
single-electron transistor}


\author{Yasuhiro Utsumi\corauthref{cor1}},
\ead{utsumi@cmt.is.tohoku.ac.jp}
\author{Hiroshi Imamura}, 
\author{Masahiko Hayashi}, 
\author{Hiromichi Ebisawa}
\corauth[cor1]{Corresponding author. Fax: +81-22-229-5051}

\address{Graduate School 
of Information Sciences,
Tohoku University, Sendai 980-8579, Japan
}

\begin{abstract}
% Text of abstract
We theoretically investigate quantum fluctuation of charge between even and odd states of a normal-superconducting-normal single-electron tunneling transistor. 
It is shown that due to the superconducting gap, the charge fluctuation in the  Coulomb blockade regime for even state is larger than that for odd state. 
We show that large energy correction in the former regime caused by charge fluctuation can be explained by considering the charging energy renormalization. 
\end{abstract}
\end{frontmatter}

\newcommand{\ve}{\varepsilon}
\newcommand{\vp}{\varphi}
\newcommand{\pdt}{\partial_{t}}
\newcommand{\ri}{i}
\newcommand{\rd}{{\rm d}}

% main text
%\section{Introduction}
The charge fluctuation caused by strong tunneling is one of the basic problems in physics of single-electron tunneling (SET) transistors. 
Recently, there has been much development in the theoretical research\cite{Falci,Schoeller,Golubev}. 
Schoeller {\it et al.} calculated the current by using a systematic diagrammatic technique based on the Keldysh formalism and predicted the renormalization of 
conductance and charging energy. 
%The renormalization factor at the degeneracy point is given by $1/(1+2 \alpha_0 \ln(E_C / {\rm max} \{ 2 \pi k_{\rm B} T, |eV|/2 \}))$\cite{Schoeller}, where $\alpha_0$ is the dimensionless junction conductance and $E_C$ is the charging energy. 
The conductance and the charging energy decrease with decreasing temperature or bias voltage. 
Recently, Joyez {\it et al.} observed the conductance renormalization at low temperature\cite{Joyez}. 

As for superconducting islands, the parity effect\cite{RBT,Utsumi1} has been observed. 
The parity effect is an evidence that a Cooper pair is composed of two
electrons. 
Though there is an investigation on the large charge fluctuation which takes account of the parity effect\cite{Neumann}, there is no study which focuses on the quantum fluctuation of charge between even and odd states. 

In this paper, we mainly study the ground-state properties of a normal-superconducting-normal (NSN) SET transistor with a special attention to the charge fluctuation between even and odd states. 
This paper is complementary to a previous paper\cite{Utsumi2} where we 
studied transport properties. 
We assume that the charging energy $E_C$ exceeds the superconducting gap $\Delta$. 
We show that the charge fluctuation in the Coulomb blockade (CB) regime for even state is larger than that for odd state. 
The large energy correction caused by large charge fluctuation in CB regime for even state can be explained by considering the charging energy renormalization. 
%\section{Model and Calculations}

%
%Figure1
%===========================================================
%
%Figure1
\begin{figure}[ht]
\begin{center}
\includegraphics[width=1\linewidth]{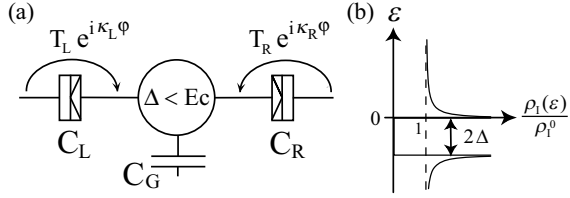}
\caption{
(a) The equivalent circuit of a NSN transistor. 
(b) The density of state when there are odd number of electrons in the superconducting island. 
}
\label{fig:system}
\end{center}
\end{figure}
%===========================================================

Figure \ref{fig:system} (a) shows the equivalent circuit of the NSN transistor. 
A superconducting island exchanges quasiparticles (QPs) with a left (right) lead via a tunnel junction characterized by the tunneling matrix element $T_{\rm L(R)} $ and is coupled to a left (right) lead and a gate via capacitors $C_{\rm L(R)}$ and $C_{\rm G}$. 
In the following, we consider the symmetric system where $T_{\rm L}=T_{\rm R}$ and  $C_{\rm L}=C_{\rm R}$. 
We limit ourselves to the discussion on charge fluctuation at zero temperature where the lowest two energy states make a dominant contribution. 
In the language of effective spin $\sigma$, the total Hamiltonian is written as
%^^^^^^^^^^^^^^^^^^^^^^^^^^^^^^^^^^^^^^^^^^^^^^^^^^^^^^^^^^^
\begin{eqnarray}
H 
&=&
\sum_{{\rm r}, k, n}
\ve_{{\rm r} k}
\:
a_{{\rm r} k n}^{\dag} 
a_{{\rm r} k n}
+
\sum_{\stackrel{\scriptstyle{\rm r=L,R}}{k, k', n}}
(
T_{\rm r} e^{\ri \kappa_{\rm r} \varphi}
\:
a_{{\rm I} k n}^{\dag} 
a_{{\rm r} k' n}
\sigma_{+}
\nonumber
\\
&+&
{\rm h. c.})
+
\Delta_0 (\sigma_{z}+1)/2+E_0,
\label{eq:capacitance_model}
\end{eqnarray}
%^^^^^^^^^^^^^^^^^^^^^^^^^^^^^^^^^^^^^^^^^^^^^^^^^^^^^^^^^^^
where the first term is the Hamiltonian of QPs in the lead r (=L,R) and the island (r=I). 
$a_{{\rm r} k n}$ is the annihilation operator of a QP with wave vector $k$, transverse channel $n$ which includes the spin. 
We consider the limit of large number of transfer channels, $N_{\rm ch}$. 
We assume that an unpaired electron occupies the lowest energy level above the superconducting gap. 
Figure \ref{fig:system}(b) shows the normalized density of state (DOS) in the island $\rho_{\rm I}(\ve)/\rho^0_{\rm I}$. 
The second term describes the tunneling across junctions and simultaneous change of the charge state. 
$\varphi$ is the phase difference between the left and right leads and 
$\kappa_{\rm L}=-\kappa_{\rm R}=1/2$. 
The third term describes the excitation energy, namely the energy difference between two charge states, which is written with a gate charge $Q_{\rm G}$ as $\Delta_0=E_C(1-2 Q_{\rm G}/e)+\Delta$. 
The last term $E_0=E_C (Q_G/e)^2$ is the charging energy for the charge state $\left|0\right>$ where there are even number of electrons in the island. 

We employ a mapping of the spin-1/2 operator onto two fermion operators, $c$ and $d$, \cite{Isawa} and use the Schwinger-Keldysh approach\cite{Utsumi2,Chou}. 
The approximate generating functional is expressed by 
%----------------------------------------------------------
$
\bar{W} = - \ri \hbar \ln {\rm Tr}[G_c^{-1}],
$
%----------------------------------------------------------
where trace is performed over the time $t$ defined on the closed time-path $C$ which goes from $-\infty$ to $\infty$, returns along the real axis and closes at $-\infty-\ri \hbar \beta$. 
Here $G_c$ is the full Green function of $c$ field defined as 
$G_c^{-1}(t_1,t_2)
=
(\ri \hbar \pdt-h(t_1))\delta(t_1,t_2)-\Sigma_c(t_1,t_2)$, 
where $\delta$ is $\delta$-function defined on $C$ and we replace $\Delta_0$ with $h(t)$ in order to preform the functional derivative. 
$h(t)$ is regarded as the \lq\lq external magnetic field" acting on the spin $\sigma$. 
The $c$ field self-energy is given by the product of particle hole propagator and the free Green function of $d$ field: 
$\Sigma_c(t_1,t_2) = \ri \hbar g_{\phi}(t_1,t_2) \alpha(t_1,t_2)$. 

The average charge in the island is calculated by functional derivative of the generating functional $\bar{W}$ with respect to the magnetic field, 
%----------------------------------------------------------
$
Q/e-1/2
=
\left.
\delta \bar{W}/\delta h_{\Delta}(t)
\right|_{h_{\Delta}=0, h_c=\Delta_0},
$
%----------------------------------------------------------
where $h_{\Delta(c)}$ is the relative (center-of-mass) coordinate\cite{Chou} of the magnetic field. 
The phase difference is fixed as $\vp_c(t)=eVt/\hbar$ and $\vp_{\Delta}(t)=0$. 
The average charge is rewritten as, 
%----------------------------------------------------------
\begin{eqnarray}
Q/e
&=&
\frac{1}{2}+
\int \frac{\rd \ve}{\pi}
\frac{\alpha^R(\ve)}
{\alpha^K(\ve)}
{\rm Im} G_c^{R}(\ve),
\label{eq:averagecharge}
\end{eqnarray}
%----------------------------------------------------------
where 
$\alpha^K(\ve)=-2 \pi \ri \sum_{\rm r=L,R} 
\alpha_{\rm r}^0 |\rho(\ve-\Delta-\kappa_{\rm r} eV)|$ 
and 
$\alpha^R(\ve)=-\pi \ri \sum_{\rm r=L,R} 
\alpha_{\rm r}^0 \rho(\ve-\Delta-\kappa_{\rm r} eV)$ 
are the Keldysh and retarded component of the particle-hole propagator, respectively. 
Here 
$\alpha^0_{\rm r}
=
N_{\rm ch} T^2 \rho_{\rm I} \rho_{\rm r}$
and $\rho_{\rm r}$ is DOS in the lead ${\rm r}$. 
The spectral density of $\rho(\ve)$ is suppressed around $\ve=0$ due to the superconducting gap\cite{Utsumi2}. 

In Eq. (\ref{eq:averagecharge}), the imaginary part of the retarded full $c$ field Green function $G_c^R(\ve)=1/(\ve-\Delta_0-\Sigma_c^R(\ve))$ represents the spectral function of the excitation of the charge state\cite{Utsumi2}. 
${\rm Im} [\Sigma_c^R]=-\gamma$ is the life-time broadening caused by dissipative charge fluctuation. 
The real part of the self-energy is the shift of a pole of the Green function. 
It is noticed that at zero-bias voltage Eq. (\ref{eq:averagecharge}) is formally equivalent to the expression obtained within the resonant tunneling approximation for normal metal island\cite{Schoeller}. 

%\section{Result and Discussions}
%%%%%%%%%%%%%%%%%%%%%%%%%%%%%%

%Figure2
%===========================================================
%
%Figure2
\begin{figure}[ht]
\begin{center}
\includegraphics[width=1\linewidth]{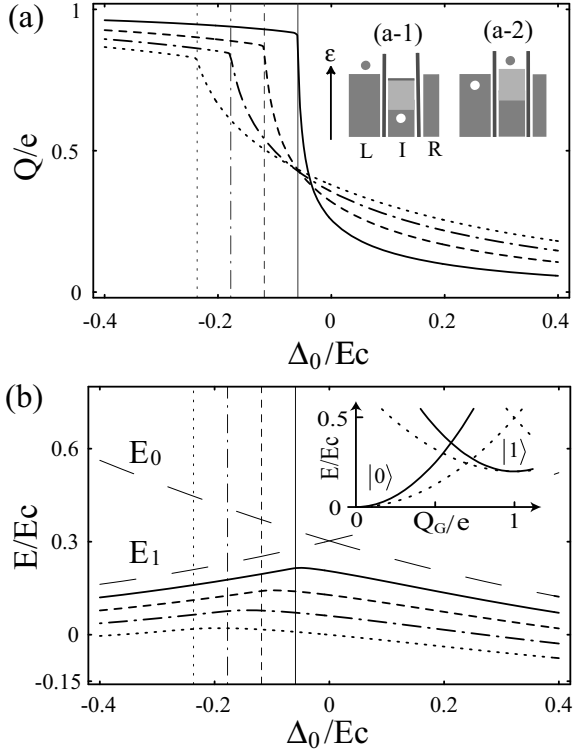}
%
%\end{center}
%\end{figure}
%\begin{figure}[ht]
%\begin{center}
%
\caption{
(a) The average charge as a function of the excitation energy at zero-bias voltage for $\alpha_0=0.1$ (solid line), 0.2 (dashed line) 0.3 (dot-dashed line) and 0.4 (dotted line). 
Vertical lines indicate positions of degeneracy points\cite{Utsumi1}. 
Two insets show schematic diagrams of dominant process of charge fluctuation in CB regime for odd state (panel (a-1)) and that for even state (panel (a-2)). 
In the panel (a-1)/(a-2), the island chemical potential is under/above that of leads. 
In CB regime for odd state, the charge fluctuation is blocked due to the superconducting gap. 
(b) The ground-state energy as a function of the excitation energy for various $\alpha_0$. 
The long dashed lines indicate the charging energy of even ($E_0$) and odd ($E_1$) state, respectively. 
As $\alpha_0$ increases, the local maximum point shifts leftwards. 
The inset of the panel (b) shows a schematic diagram of gate charge dependence of the charging energy. 
Solid and dashed lines indicate the energy with and without charge fluctuation, respectively. 
The energy correction in CB regime for even state is larger than that for odd state. 
Parameters are chosen as $\Delta/E_C=0.1$ and $\rho_{\rm I}^0 E_C=10^{-3}$. 
The value $\Delta \rho_{\rm I}^0=10^2$ is the same order as an Al island whose volume is $10^4 {\rm nm}^3$. 
}
\label{fig:averatgenumber1}
\end{center}
\end{figure}
%===========================================================

Figure \ref{fig:averatgenumber1}(a) shows the average charge as a function of the excitation energy at zero-bias voltage for various dimensionless conductance 
$\alpha_0=\sum_{\rm r=L,R} \alpha^0_{\rm r}$. 
Vertical lines indicate positions of degeneracy points. 
Each position is given by the real part of the pole of the full $c$ field Green function. 
We can see the degeneracy point is shifted leftwards as $\alpha_0$ increases. 
The deviation of the average charge from the discretized value, in CB regime for even state is larger than that for odd state. 
This asymmetric behavior around the degeneracy point can be understood by considering processes of charge fluctuation. 
Two insets show the schematic diagrams of the dominant process of quantum fluctuation of QP in CB regime for odd (Fig. \ref{fig:averatgenumber1}(a-1)) and even (Fig. \ref{fig:averatgenumber1}(a-2)) states. 
In the former regime, the dominant process corresponds to a particle-hole excitation with a particle in one lead and a hole under the superconducting gap in the island (Fig. \ref{fig:averatgenumber1}(a-1)). 
In this case, a low-energy particle-hole excitation, which makes a large 
contribution to the charge fluctuation, is blocked due to the superconducting gap. 
On the other hand, in CB regime for even state, the dominant process corresponds to a particle-hole excitation with a hole in one lead and particle in the island (Fig. \ref{fig:averatgenumber1}(a-2)). 
Because a low-energy particle-hole excitation is enhanced due to the large DOS at the edge of the superconducting gap, the charge fluctuation is large. 

The excitation energy dependence of the ground-state energy for various $\alpha_0$ is plotted in Fig. \ref{fig:averatgenumber1}(b). 
The ground-state energy is expressed as 
%----------------------------------------------------------
\begin{eqnarray}
E
&=&
E_0
-
\int^0_{-\infty}
\rd \ve
\frac{1}{\pi}
{\rm Im}
\ln
\left( G^R_c(\ve) \right)
\label{eq:self-energy}
\end{eqnarray}
%----------------------------------------------------------
The energy correction in CB regime for even state is larger than that for odd state, because the charge fluctuation is large in the former regime. 
As $\alpha_0$ increases, the local maximum point shifts leftwards and the degeneracy point shifts in the same direction. 
This behavior can be explained by considering the charging energy renormalization\cite{Utsumi2}. 
The inset of Fig. \ref{fig:averatgenumber1}(b) shows a schematic diagram of gate charge dependence of the charging energy. 
Solid and dashed lines indicate the charging energy with and without charge fluctuation, respectively. 
One can see that the energy correction caused by the charging energy renormalization in CB regime for even state is larger than that for odd state. 

%Figure3
%===========================================================
%
%Figure3
\begin{figure}[ht]
\begin{center}
\includegraphics[width=1\linewidth]{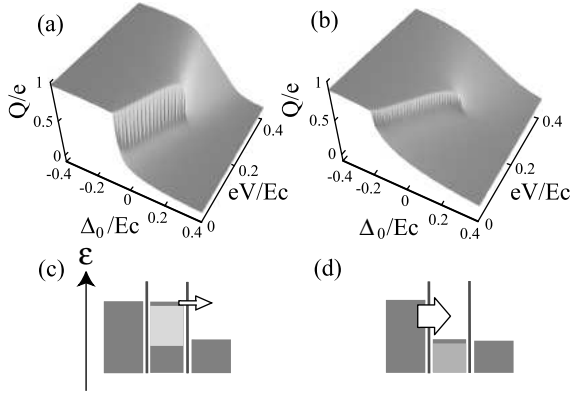}
\caption{
The 3D plot of the average charge versus the excitation energy and the bias voltage for (a) $\alpha_0=0.1$ and (b) 0.3. 
The kink structure is seen at the boundary of CB regime for even state. 
The schematic diagrams of dissipative charge fluctuation process at the boundary of CB regime for odd state (c) and that for even state (d). 
In the panel (c)/(d), the island chemical potential matches that of the left/right lead. 
(c) The fluctuation is blocked by the superconducting gap. 
(d) The fluctuation increases with increasing bias voltage. }
\label{fig:averagecharge2}
\end{center}
\end{figure}
%===========================================================
The average charge at finite bias voltage is shown in Figs. \ref{fig:averagecharge2}(a) and  \ref{fig:averagecharge2}(b). 
Even for large $\alpha_0$ (Fig. \ref{fig:averagecharge2}(b)), there is kink 
structure at the boundary of CB regime for even state. 
The kink structure is robust against charge fluctuation in the range of $eV < 2 \Delta$, because dissipative charge fluctuation is blocked due to the superconducting gap (Fig. \ref{fig:averagecharge2} (c)), and only the unpaired electron contributes to the life-time broadening: $\gamma \sim \pi \alpha_0 /2 \rho_{\rm I}^0$. 
On the other hand at the boundary of CB regime for odd state, the dissipative charge fluctuation increases with increasing bias voltage 
(Fig. \ref{fig:averagecharge2} (d)), and the life-time 
broadening $\gamma$ is proportional to the number of QP states between the chemical potential of the left lead and that of the island:
$\gamma \sim \pi \alpha_0 \sqrt{\Delta \ eV}$. 
Therefore, a small variation of the average charge at the boundary is smeared by applied bias voltage. 

%\section{Conclusion}

In conclusion, we have theoretically investigated the charge fluctuation between even and odd states of the superconducting island. 
The charge fluctuation is suppressed due to the superconducting gap in CB regime for odd state, while it is enhanced in CB regime for even state due to the large DOS at the superconducting gap edge. 
The large charge fluctuation causes the large energy correction in CB regime for even state, which can be understood by considering the charging energy renormalization. 
At finite bias voltage, there is the kink structure in the 3D plot of the average charge at the boundary of CB regime for even state, even for large $\alpha_0$. 
This is because the dissipative charge fluctuation is suppressed due to the superconducting gap. 

We would like to thank Y. Isawa, Y. Takane and A. Kanda for valuable discussions.  
One of us (H. Imamura) is supported by the Ministry of Education, Culture, Sports, Science and Technology, Grant-in-Aid for Encouragement of Young Scientists, 13740197, 2001.

\clearpage
%***********************************************************
\end{document}